\documentclass[aps,pra,twocolumn,superscriptaddress]{revtex4-2}

\usepackage{braket}
\usepackage{graphicx}
\usepackage{amsmath,verbatim,latexsym,amssymb,indentfirst,mathrsfs,mathtools,amsthm,bbm,bm,hyperref,url}

\newcommand{\tr}{\mbox{Tr}}

\usepackage{color}
\usepackage{threeparttable}
\usepackage{booktabs}
\usepackage{array}
\usepackage{siunitx}
\usepackage{makecell}
\usepackage[dvipsnames]{xcolor}
\usepackage[normalem]{ulem}

\newcommand{\daria}[1]{{\color{black}#1}}
\newcommand{\kater}[1]{{\color{black}#1}}

\usepackage[title,titletoc]{appendix}

\makeatletter
\def\p@subsection{}
\makeatother
\makeatletter
\def\p@subsubsection{}
\makeatother

\usepackage{nameref}
\usepackage{zref-xr}
\zxrsetup{toltxlabel}
\zexternaldocument*{sm}

\begin{document}
	\title{Entanglement assisted probe of the non-Markovian to Markovian transition in open quantum system dynamics}
	
	
 \author{Chandrashekhar Gaikwad}\thanks{Equal contribution}
 \affiliation{Department of Physics, Washington University, St. Louis, Missouri 63130, USA}
  \author{Daria Kowsari}\thanks{Equal contribution}
  \affiliation{Department of Physics, Washington University, St. Louis, Missouri 63130, USA}
  \affiliation{Center for Quantum Information Science and Technology, University of Southern California, Los Angeles, California 90089, USA}
\affiliation{Department of Physics \& Astronomy, University of Southern California, Los Angeles, California 90089, USA}
  \author{Carson Brame}
    \affiliation{Department of Physics, Washington University, St. Louis, Missouri 63130, USA}
    \author{Xingrui Song}
        \affiliation{Department of Physics, Washington University, St. Louis, Missouri 63130, USA}
        \author{Haimeng Zhang}
\affiliation{Center for Quantum Information Science and Technology, University of Southern California, Los Angeles, California 90089, USA}
\affiliation{Ming Hsieh Department of Electrical \& Computer Engineering, University of Southern California, Los Angeles, California 90089, USA}
 \author{Martina Esposito}
 \address{CNR-SPIN Complesso di Monte S. Angelo, via Cintia, Napoli 80126, Italy}
\author{Arpit Ranadive}
\affiliation{Universit\'e Grenoble Alpes, CNRS, Grenoble INP, Institut N\'eel, 38000 Grenoble, France}
\author{Giulio Cappelli}
\affiliation{Universit\'e Grenoble Alpes, CNRS, Grenoble INP, Institut N\'eel, 38000 Grenoble, France}
\author{Nicolas Roch}
\affiliation{Universit\'e Grenoble Alpes, CNRS, Grenoble INP, Institut N\'eel, 38000 Grenoble, France}
\author{Eli M. Levenson-Falk}
\affiliation{Center for Quantum Information Science and Technology, University of Southern California, Los Angeles, California 90089, USA}
\affiliation{Department of Physics \& Astronomy, University of Southern California, Los Angeles, California 90089, USA}
\affiliation{Ming Hsieh Department of Electrical \& Computer Engineering, University of Southern California, Los Angeles, California 90089, USA}
	\author{Kater W. Murch}
	\email[]{murch@physics.wustl.edu}
	\affiliation{Department of Physics, Washington University, St. Louis, Missouri 63130, USA}

	\begin{abstract}
We utilize a superconducting qubit processor to experimentally probe \kater{non-Markovian dynamics of an entangled qubit pair.}  We prepare an entangled state between two qubits and monitor the evolution of entanglement over time as one of the qubits interacts with a small quantum environment consisting of an auxiliary transmon qubit coupled to its readout cavity. We observe the collapse and revival of the entanglement as a signature of quantum memory effects in the environment.  We then engineer the non-Markovianity of the environment by populating its readout cavity with thermal photons to show a transition \kater{from non-Markovian to Markovian dynamics, ultimately} reaching a regime where the quantum Zeno effect creates a decoherence-free subspace that effectively stabilizes the entanglement between the qubits. 
\end{abstract}
	
\date{\today}
\maketitle

Decoherence is a ubiquitous challenge in quantum technologies. At a microscopic level, decoherence arises from the entanglement of a quantum system with degrees of freedom in its environment. Without access to these degrees of freedom, information about the quantum state is lost \cite{Zurek2003, Jacquod2009}. The monotonic reduction in a quantum state's coherence is typically described by the well-known Gorini-Kossakowski-Sudarshan-Lindblad (GKSL) master equation \cite{Lindblad1976, Gorini1976} for the system's density operator $\rho$. In particular, the GKSL master equation is valid under the Born-Markov set of approximations, which assume both weak coupling to the environment, and that the environment is Markovian, i.e.~memoryless \cite{Breuer2007}. This mathematically amenable description is surprisingly effective in describing a broad range of quantum dynamics. Moreover, in the Markovian regime, dissipation engineering by an intentional introduction of Markovian dissipation has been employed as a powerful method of quantum control; with applications including error correction \cite{Kraglund2020, Leghtas2013, Touzard2018}, state preparation \cite{Murch2012,Magnard2018}, state stabilization \cite{Holland2015, Schwartz2016}, and quantum simulation \cite{harrington_review}. Naturally, however, there is another paradigm of decoherence known as the non-Markovian regime, where quantum memory effects induced by large system-environment correlations thwart a Markovian description. In this regime, the dynamics of the system is governed by the generalized Nakajima-Zwanzig master equation \cite{Nakajima1958, Zwanzig1960} which incorporates the memory effects of the environment.

Non-Markovian dynamics have the potential to enable novel applications stemming from memory effects in the environment, such as new approaches towards fault-tolerant quantum computation \cite{Aharonov2006, Alicki2006, Rossini2023}, \kater{quantum control \cite{Reich2015}}, fidelity improvement in the implementation of the teleportation algorithms \cite{Liu2020}, and coherence preservation \cite{Vlachos2022}. \kater{The non-Markovianity of an open quantum system can be measured using two common methods \cite{Wolf2008}.} The most prominent measure is known as the \emph{trace distance} method proposed by Breuer et.\ al.\ \cite{Breuer2009}, which relies on the fact that any completely positive trace-preserving (CPTP) quantum map between two-time steps will only result in a decrease of the distinguishability between two quantum states, hence any increase in the distance between states is associated with memory effects \cite{Breuer2016}. Later, an \emph{entanglement measure} was introduced by Rivas et.\ al.\ \cite{Rivas2010}, where one probes quantum memory effects by allowing part of an entangled pair to interact with an environment. Again, a CPTP map will only decrease the degree of entanglement and an increase in entanglement during the system evolution is a signature of quantum memory effects. Both methods \cite{Haikka11} have been employed to observe signatures of non-Markovianity, notably in \kater{nitrogen-vacancy centers \cite{Dong2018, Haase2018, Wang2018a,PhysRevLett.124.210502}, photonic systems \cite{PhysRevLett.104.100502,Liu2011, Liu2013, Bernardes2015, Wu2020}, nuclear magnetic resonance \cite{PhysRevA.99.022107, Wu2020, Chen2022}, trapped ions \cite{Gessner2014}, and on superconducting processors \cite{White2020}}.

In this Letter, we harness the entanglement between two superconducting qubits as a probe of quantum memory effects.  We initialize the qubits in a Bell state and study the qubits' concurrence \cite{Wooters1998} over time as one of the qubits interacts with a small quantum environment consisting of a third qubit dispersively coupled to a microwave resonator. We observe collapse and then revival of the qubits' concurrence as a clear signature of the non-Markovian nature of the environment as the qubit becomes entangled and then disentangled with the environment.  The non-Markovianity of the environment is then tuned by introducing Lindblad dephasing \kater{on the environment \cite{Pang2017}}. This allows us to investigate \kater{a transition away from non-Markovian dynamics to a regime where the GKSL master equation describes the dynamics. Since the GKSL master equation requires a Markovian approximation, we refer to this regime as the Markovian regime}. In this \kater{Markovian} regime, we further increase the dissipation on the environment, ultimately reaching a regime where the quantum Zeno effect pins the environment state, thereby preserving the qubits' entanglement. 

	\begin{figure}
		\centering
		\includegraphics[width=0.5\textwidth]{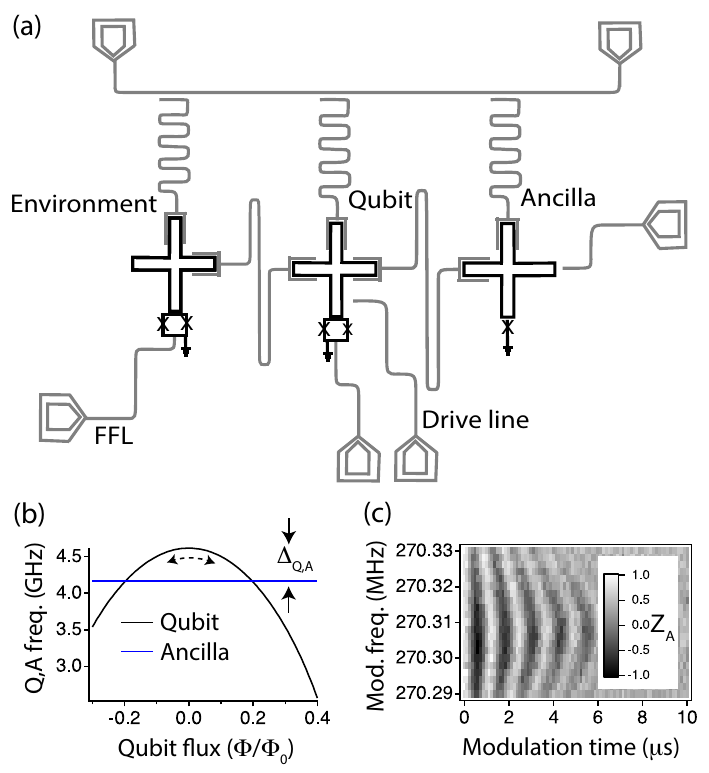}
		\caption{{\bf Experiment setup.} (a) Sketch of the experiment which includes three qubits respectively labeled ``Environment", ``Qubit", ``Ancilla". The qubits share resonators that mediate nearest-neighbor coupling. Each qubit is coupled to a readout resonator, which can be probed by a common feedline.  The Environment and Qubit are frequency tunable via on-chip fast flux lines (FFL). (b) The respective frequencies of the Qubit  and Ancilla; resonant coupling between the qubits is achieved by applying a parametric modulation of the Qubit at roughly $\Delta_\mathrm{Q,A}/2$. (c) When the Qubit is prepared in its excited state, parametric resonance can be observed by examining the Ancilla excitation versus modulation frequency. }
		\label{setup}
	\end{figure}

 Figure \ref{setup}(a) displays the basic setup of the experiment, \kater{with the system Hamiltonian given in \cite{Suppl_Mat}}. The experiment comprises a three-qubit processor with individual readout resonators dispersively coupled to each qubit and nearest-neighbor qubits sharing a resonator mediated coupling. 
 The readout resonators allow us to perform individual state readouts of the three qubits by probing the associated microwave resonators with a microwave drive. We first focus on a sub-portion of the processor with two qubits denoted as the ``Qubit'' and the ``Ancilla". The Qubit is frequency tunable via a Superconducting QUantum Interference Device (SQUID) loop and the Ancilla is fixed-frequency, both designed to be in the transmon regime \cite{Koch2007}. In order to minimize the decoherence effects from flux noise, we operate the Qubit at its flux sweet spot and introduce coupling to the Ancilla via parametric modulation \cite{Reagor2018}. To this end, we apply an ac radio frequency drive on the Qubit fast flux line at roughly half the detuning between the Qubit and Ancilla (Fig.~\ref{setup}(b)). We identify the resonance condition between the Qubit and Ancilla by initializing the qubit in its excited state and then applying the parametric modulation for a variable duration.  Figure \ref{setup}(c) shows the time evolution of the Ancilla $\langle \sigma_z\rangle\equiv Z_A$ near the parametric resonance. We observe a clear chevron profile with detuning from which we extract a parametric coupling rate of $\Omega_\mathrm{Q,A}/2\pi = 0.477$ MHz.

 	\begin{figure}
		\centering
		\includegraphics[width=0.5\textwidth]{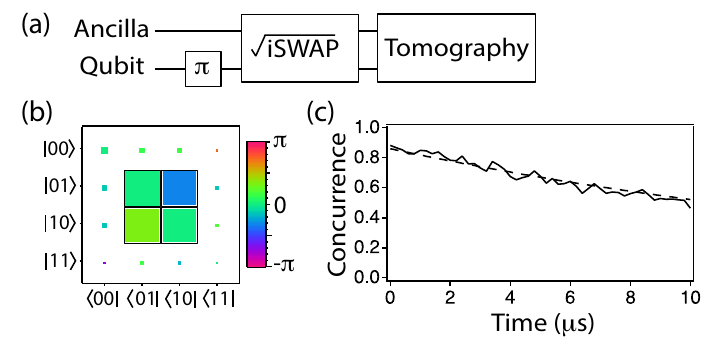}
		\caption{{\bf Qubit-Ancilla entanglement.} (a) We prepare an entangled state by initializing the Qubit in the excited state, and then applying a $\sqrt{i\mathrm{SWAP}}$ gate via parametric modulation.  (b) Quantum state tomography allows us to reconstruct the Qubit--Ancilla density operator, yielding an entangled state of the form $\frac{1}{\sqrt{2}} (\ket{10}+ e^{i\phi} \ket{01})$. (c) The measured Qubit--Ancilla concurrence versus time (solid line); we observe a monotonic decrease in the entanglement over time consistent with the single-qubit dephasing rates (dashed line).}
		\label{QAentanglement}
	\end{figure}
	
 We utilize this parametric coupling to produce a Bell state between the Qubit and Ancilla, as depicted in Fig.~\ref{QAentanglement}(a). After applying a $\pi$ rotation to the Qubit, we activate the parametric coupling for $530$ ns, corresponding to a $\sqrt{i\mathrm{SWAP}}$ gate, in principle, leaving the Qubit and Ancilla in a state, $\frac{1}{\sqrt{2}} (\ket{10}+ e^{i\phi} \ket{01})$ \cite{Caldwell2018}. We utilize quantum state tomography of the Qubit and Ancilla to characterize the resulting entangled state. For this, we measure $9$ Pauli expectation value pairs, $\{\langle \Sigma_\mathrm{Q} \Sigma_\mathrm{A}\rangle\} $, with $\Sigma_\mathrm{Q,A} \in \{X,Y,Z\}$ by simultaneously measuring the state of both the Qubit and Ancilla \cite{Kundu2019}. The average readout fidelities of the Qubit and Ancilla are respectively $0.97$ and $0.96$. As discussed in the Supplementary Materials, we use a maximum likelihood estimation method \cite{James2001} to determine the components of the Qubit--Ancilla density matrix, displayed in Fig.~\ref{QAentanglement}(b). We observe a Bell state fidelity of $0.91$, corresponding to a concurrence of $0.89$.

 With the Qubit and Ancilla entangled, we now study the evolution of the entanglement over time as the system sits idle. We display the Qubit--Ancilla concurrence versus time in Fig.~\ref{QAentanglement}(c). \kater{The concurrence slowly decreases over a timescale consistent with the respective individual dephasing times of the Qubit ($T_2^{*(Q)} = 39\ \mu$s) and Ancilla ($T_2^{*(A)} = 41\ \mu$s), e.g. $\mathcal{C}\propto \exp(-t/T_2^{*(Q)} - t/T_2^{*(A)})$, as given by the dashed line in Fig.~\ref{QAentanglement}(c).}

	\begin{figure}
		\centering
		\includegraphics[width=0.5\textwidth]{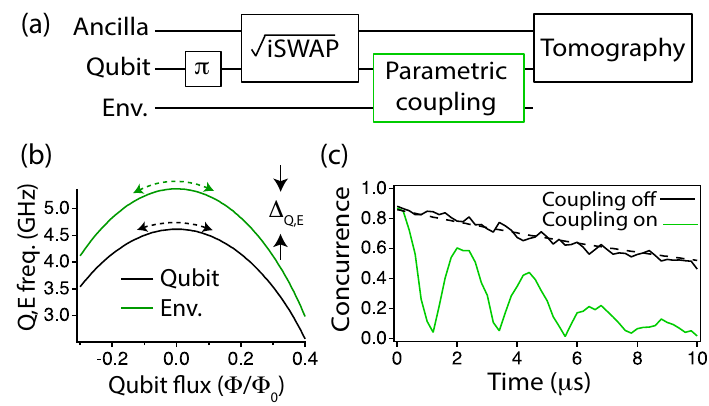}
		\caption{{\bf Concurrence revival due to non-Markovianity.}  (a) We prepare an entangled state between the Qubit and Ancilla, then we apply a parametric coupling between the Qubit and Environment, 
  finally, we perform a set of tomography pulses to reconstruct the density matrix of the Qubit--Ancilla subspace. The concurrence evolution is realized by varying the length of the Qubit--Environment parametric coupling pulses. (b) The respective frequencies of Qubit and Environment showing the Qubit--Environment detuning. (c) Concurrence evolution as a function of the Qubit--Environment parametric coupling pulse length (green). The black curve shows the concurrence evolution when the parametric coupling is turned off. 
		}
		\label{nonmark}
	\end{figure}

We now turn to studying the interaction of the Qubit--Ancilla subspace with the Environment. As displayed in Fig.~\ref{nonmark}(a), after preparing the Qubit--Ancilla in an initial Bell state, we introduce a parametric coupling between the Qubit and Environment. In this case, we apply flux modulation simultaneously to both the Qubit and Environment (Fig.~\ref{nonmark}(b)) bringing the two into parametric resonance. Both Qubit and Environment  are modulated at approximately one-quarter of their detuning ($~\Delta_\mathrm{Q,E}/4 = 2\pi\times 175\ \mathrm{rad.}/\mu\mathrm{s}$), which introduces a resonant transverse coupling between the Qubit--Environment pair at a rate of $\Omega_\mathrm{Q,E} = 2\pi\times 0.473\ \mathrm{rad.}/\mu\mathrm{s}$, limited by the resonator-mediated coupling between the pair.  After applying the parametric coupling between Qubit and Environment, we perform quantum state tomography on the Qubit--Ancilla subsystem to determine the remaining concurrence.

Figure~\ref{nonmark}(c) displays the evolution of the concurrence when the interaction between the Qubit and Environment is introduced.  In comparison to the monotonic decrease in entanglement observed previously (black curve), we now note a rapid decrease in entanglement, with clear revivals at later times (green curve). The initial decrease is expected from the principle of monogamy of entanglement \cite{Coffman_00}. Since the Qubit--Ancilla system is in a maximally entangled state, the entanglement between the Qubit--Environment introduced by the parametric coupling must cause the Qubit--Ancilla entanglement to decrease. The revival of entanglement occurs as the Qubit--Environment coupling continues and the Environment state is swapped back into the Qubit. This revival of entanglement is a clear indicator of non-Markovianity, indicating that the environment has quantum coherent memory. This is indeed expected since the environment is itself a simple two-level system. The non-Markovianity of the system can be can be calculated as \cite{Rivas2010},
%
\begin{align}\label{eq:conc_meas}
\mathcal{N} = \int_{t_0}^{t_f} dt \, \bigg|\frac{d\mathcal{C}[\rho_\mathrm{Q,A} (t)]}{dt}\bigg| - \Delta \mathcal{C},
\end{align}
where $\mathcal{C}[\cdot]$ denotes the concurrence measure, $\Delta \mathcal{C}$ is the difference in the concurrence at the initial and final steps of the evolution, and $\rho_\mathrm{QA}$ represents the Qubit--Ancilla density matrix. To elaborate, we look at the time derivative of the concurrence over the entire time evolution of the system $\in [t_0=0\ \mu\mathrm{s}, t_f=10\ \mu\mathrm{s}]$ at discrete time steps. It is clear from Eq.~\ref{eq:conc_meas} that the positive slope of the concurrence contributes to the non-Markovianity measure. By applying Eq.~\ref{eq:conc_meas} to the data in Fig.~\ref{nonmark}(c), we achieve a non-Markovianity of $\mathcal{N}=1.4$.

	\begin{figure}
	\centering
		\includegraphics[width=0.5\textwidth]{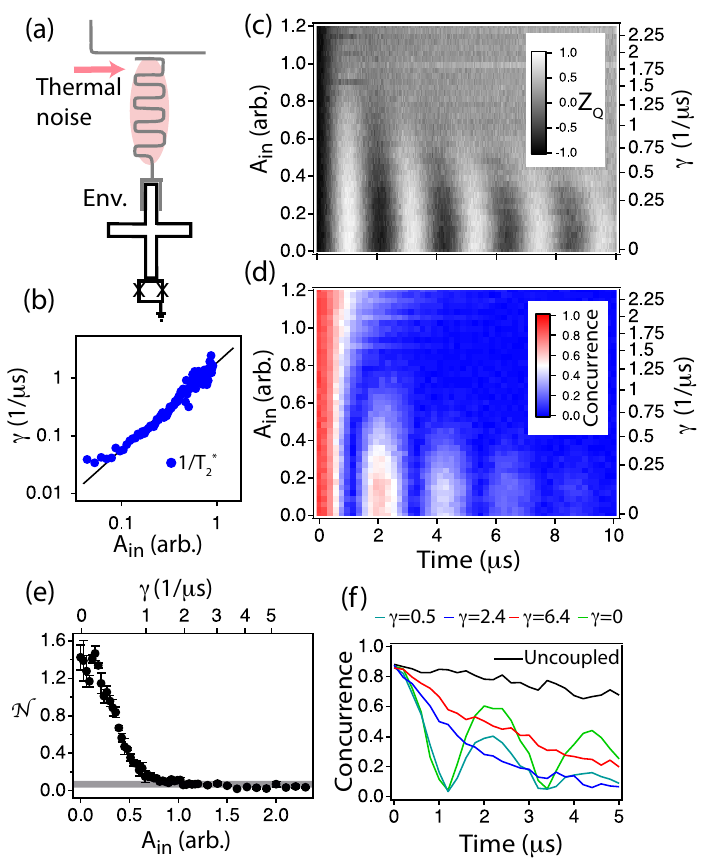}
		\caption{{\bf Non-Markovian to Markovian transition.} (a) By driving the Environment's readout resonator with pseudo-thermal noise of amplitude $A_\mathrm{in}$ we tune the Environment's memory. (b) This memory is quantified through Ramsey measurements on the Environment to determine the dephasing rate $\gamma$ versus $A_\mathrm{in}$. (c) For each value of $A_\mathrm{in}$ we calibrate the frequency of the parametric drive between the Qubit and Environment by studying $Z_Q$ versus time and maximizing the population transfer \cite{Suppl_Mat}. (d) The Qubit--Ancilla concurrence versus time for different Environment dephasing rates. The transition to monotonic behavior indicates the transition from non-Markovian to Markovian dynamics.  (e) The non-Markovian measure (\ref{eq:conc_meas}) quantified across the transition. The error bars indicate the standard error of the mean from three independent experimental trials. The gray bar indicates the measure applied to the case where the Environment is decoupled and characterizes the background of the measure. (f) The concurrence versus time for a few specific dephasing rates (expressed in units of $\mu\mathrm{s}^{-1}$).} 
        \label{transition}
	\end{figure}

With a clear demonstration of non-Markovian dynamics, we now study how this measure changes as the memory of the Environment is tuned.  We achieve this by expanding the size of the Environment to include the quantum states of light that occupy the microwave resonator that is dispersively coupled to the Environment.  So far, we have considered this resonator to remain in the vacuum state, which does not affect the Environment's memory. Now, we introduce pseudo-thermal photons into this resonator via a noisy microwave drive as indicated in Fig.~\ref{transition}(a). The interaction between the Environment and its resonator is captured by the simple dispersive coupling Hamiltonian, $H_\mathrm{int} = \chi a^\dagger a \sigma_z^\mathrm{E}$, where $\chi/2\pi=200~\mathrm{kHz}$ is the dispersive coupling rate, $a^\dagger a$ is the resonator photon number, and $\sigma_z^\mathrm{E}$ is the Pauli operator that acts on the Environment in the energy basis.  This interaction can be viewed as either an Environment-state-dependent frequency shift on the resonator frequency, whereby photons carry away information about the state of the Environment, or as an ac-Stark shift of the qubit frequency, whereby the fluctuating intra-resonator photon number dephases the qubit \cite{hatr13,Murch2012}.  

The noisy drive on the cavity is chosen to have a bandwidth (1.8 MHz) that exceeds $\chi$, ensuring a uniform drive independent of the Environment state. Furthermore, this drive is set to have a correlation time (90 ns) much shorter than any other timescale of the dynamics, allowing us to treat its dephasing effect as Markovian. We calibrate the dephasing via direct Ramsey measurements on the Environment. This establishes a relationship between the dephasing rate and the noise amplitude ($A_\mathrm{in}$) as shown in Fig.~\ref{transition}(b). We find an empirical relationship for the Environment dephasing $\gamma = 1.84 (\mu\mathrm{s})^{-1}A_\mathrm{in}^{1.5}$, 
 as given by the black line in Fig.~\ref{transition}(b).

The introduction of the thermal photons into the Environment causes slight shifts in the parametric coupling between the Qubit and the Environment. As such, we calibrate the parametric coupling between the Qubit and Environment for each value of $A_\mathrm{in}$. Figure \ref{transition}(c) shows the resulting parametric coupling between the Qubit and Environment when the Qubit is initialized in the exited state and the parametric coupling is activated for a variable duration of time.  By increasing the dephasing of the Environment, we observe diminished population transfer contrast between the Qubit and the Environment.

Next, we investigate the time evolution of the Qubit--Ancilla concurrence for different values of the Environment dephasing. Increasing the Environment dephasing induces a transition from non-Markovian to Markovian dynamics as displayed in Fig.~\ref{transition}(d). \kater{We quantify the transition away from non-Markovian dynamics via the measure (Eq. \ref{eq:conc_meas}) as displayed in Fig.~\ref{transition}(e); as the dephasing of the Environment is increased beyond $\gamma \simeq 1\ (\mu\mathrm{s})^{-1}$, $\mathcal{N}$ becomes consistent with zero. However, the dynamics are not immediately Markovian, which we define by the applicability of the GKSL master equation to the Qubit--Ancilla subsystem. As we study in \cite{Suppl_Mat}, the GKSL master equation yields exponentially decaying dynamics of the concurrence. This matches well the measured dynamics for $\gamma \gtrsim 3\ (\mu\mathrm{s})^{-1}$, but fails to capture the dynamics for smaller values of $\gamma$. As such, the transition between these two regions, as defined, is not abrupt.}

\begin{figure}
	\centering
		\includegraphics[width=0.5\textwidth]{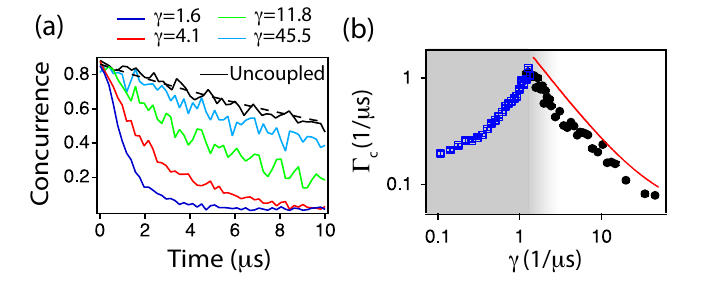}		\caption{{\bf Quantum Zeno stabilization of entanglement.} (a) Qubit--Ancilla concurrence versus time for different Environment dephasing rates; as the dephasing increases, the entanglement decay approaches the uncoupled case consistent with the Qubit and Ancilla's individual dephasing rates. (b) The exponential decay rate of the concurrence ($\Gamma_\mathrm{c})$ versus Environment dephasing rate. The gray region indicates the non-Markovian regime, where we determine $\Gamma_\mathrm{c}$ by fitting the overall (non-monotonic) decay envelope of the concurrence. In the Markovian regime, we observe that the Zeno effect suppresses the concurrence decay induced by the environment, in agreement with the expected scaling (red line). } 
        \label{zeno}
	\end{figure}

 In Fig.~\ref{transition}(f) we display the concurrence versus time for a few selected values of $\gamma$. We note two important trends; first, in the non-Markovian regime, increasing dephasing accelerates the decay envelope of the concurrence (compare $\gamma=0$ and $\gamma=0.5$), and second, in the Markovian regime, further increasing dephasing slows the decay of the concurrence ($\gamma=2.4$ and $\gamma=6.4$).  This can be understood within the context of the quantum Zeno effect \cite{Harrington2017,Misra1977,Itano1990,Kakuyanagi2015,HacohenGourgy2018,HacohenGourgy2018,Blumenthal2022,Slichter2016, PhysRevResearch.2.023133}. The thermal photons perform measurement (at rate $\gamma$) of the Environment, which slows the coupling induced by the parametric drive. In Fig.~\ref{zeno}, we explore in detail how the dephasing of the Environment affects the Qubit--Ancilla entanglement. Figure~\ref{zeno}a displays the concurrence versus time for several values of the dephasing in the Markovian regime. For increasing measurement on the Environment, we see that the decay of concurrence is slowed, approaching the limiting case where the Qubit is completely uncoupled from the Environment. We characterize the exponential decay of the concurrence with a rate $\Gamma_\mathrm{c}$, and display this rate versus Environment dephasing in Fig.~\ref{zeno}b in both the non-Markovian and Markovian regimes; the transition between these two regimes coincides with the onset of Zeno stabilization of the entanglement. Under the standard analysis of the Zeno effect \cite{PhysRevA.77.012112,Blumenthal2022}, we expect $\Gamma_\mathrm{c} = \Omega_\mathrm{Q,E}^2/4\gamma + \Gamma_0$. Here $\Gamma_0 = 1/T_2^{*Q} +1/T_2^{*A}$ is the decay rate of the concurrence when the Environment is decoupled.  We observe close agreement with this expected scaling (red curve). 
 This demonstrates a new approach to preserving quantum entanglement via Zeno-enabled pinning of environment states. 

In conclusion, we have quantified \kater{the transition from non-Markovian dynamics to Zeno dynamics with an entanglement-assisted probe.} Importantly, the probe is sensitive to the \emph{quantum memory} of the environment; a classical environment that stores populations will not result in the revival of concurrence for the entangled probe. This approach can have utility in the test of the quantum nature of decoherence channels (e.g.\ in testing models of quantum gravity \cite{AlNasrallah2023}).  Moreover, by introducing controllable dissipation on the environment we observe stabilization of the  Qubit--Ancilla subsystem, highlighting how dissipation forms a powerful tool for quantum subspace engineering \cite{harrington_review}.



\begin{acknowledgements}
\emph{Acknowledgments}---The authors are thankful for the useful discussions with Daniel Lidar, Kade Head-Marsden, Patrick Harrington, Weijian Chen, Kaiwen Zheng, Maryam Abbasi, Serra Erdamar, and Archana Kamal. This research was supported by NSF Grants No. PHY-1752844 (CAREER) and OMA-1936388, the Air Force Office of Scientific Research (AFOSR)  Multidisciplinary University Research Initiative (MURI) Award on Programmable systems with non-Hermitian quantum dynamics (Grant No. FA9550-21-1- 0202), the John Templeton Foundation, Grant No. 61835, ONR Grant Nos. N00014-21-1-2630 and N00014-21-1-2688 (YIP), Research Corp. Grant No. 27550 (Cottrell), European Union's FET Open AVaQus grant no. 899561 and Marie Sklodowska-Curie grant no. MSCA-IF-835791. Devices were fabricated and provided by the Superconducting Qubits at Lincoln Laboratory (SQUILL) Foundry at MIT Lincoln Laboratory, with funding from the Laboratory for Physical Sciences (LPS) Qubit Collaboratory.
\end{acknowledgements}


\newpage

\textcolor{white}{.}

\newpage

\begin{widetext}

\begin{center}
    
\large \bf Supplemental Information for “Entanglement assisted probe of the
non-Markovian to Markovian
transition in open quantum system
dynamics”

\end{center}

\section{System Hamiltonian}

The Hamiltonian of the three qubit system can be expressed in terms of the Pauli operators $\sigma_x$ and $\sigma_z$ for the three qubits as ($\hbar=1$),

\begin{align}
H = \sum_{i=\mathrm{A,Q,E}} \big(-\frac{1}{2}\tilde{\omega}_{\mathrm{q},i} \sigma_z^i + (\omega_{\mathrm{c},i} - \chi_{\mathrm{qc},i})a^{\dagger}_i a_i \sigma_z^i\big) + J_\mathrm{A,Q} \sigma_x^\mathrm{A} \sigma_x^\mathrm{Q} + J_\mathrm{Q,E} \sigma_x^Q \sigma_x^E.
\end{align}


The cavity creation(annihilation) operators are denoted as $a^\dagger(a)$ with $\tilde{\omega}_\mathrm{q} = \omega_\mathrm{q} + \chi_\mathrm{qc}$ representing the Lamb-shifted qubit frequency in the dispersive frame. The dispersive coupling rates $\chi_\mathrm{qc}$, qubit frequencies, and cavity frequencies are given in Table \ref{tab:sim}. The coupling rates $J_\mathrm{A,Q}$ and $J_\mathrm{Q,E}$ are tuned by application of parametric modulation.  In addition, the induced dephasing in the environment can be described by adding a simple Lindbladian term to the master equation governing the system's time evolution, which can be formulated as,
\begin{align}
\dot{\rho}(t)  = -i[H, \rho(t)] + \gamma\mathcal{L}_\mathrm{E}\rho(t),
\end{align}
\noindent where the Lindblad operator is defined as, $\mathcal{L}_\mathrm{E}\rho = (\sigma_z^\mathrm{E}\rho\sigma_z^\mathrm{E}-\rho)$, $\gamma$ denotes the induced dephasing rate in the environment, and $\rho(t)$ is the system's density matrix at time $t$.


\section{Numerical simulation of the Qubit--Ancilla dynamics in the Markovian regime}

\daria{The Hamiltonian of the Qubit--Ancilla system in the Markovian regime can be written as ($\hbar =1$),
\begin{align}
    H_{\mathrm{A},\mathrm{Q}} = -\frac{1}{2} \omega_\mathrm{A} \sigma_z^\mathrm{A} -\frac{1}{2} \omega_\mathrm{Q} \sigma_z^\mathrm{Q}.
\end{align}
This Hamiltonian does not include any coupling between the Qubit and Ancilla, and is therefore applicable after the two qubits have been initialized in an entangled state.  The dynamics of the system in presence of dephasing on both qubits is governed by a simple GKSL master equation as,
\begin{align}\label{eq:lindblad}
    \dot{\rho}(t) = -i[H_\mathrm{A,Q}, \rho (t)] + \sum_{j=\mathrm{A},\mathrm{Q}} \frac{1}{2} \left(2 C_j \rho(t) C_j^{\dagger} - \rho(t) C_j^{\dagger} C_j - C_j^{\dagger} C_j \rho(t)\right),
\end{align}
where $C_j = \sqrt{\gamma_j} \sigma_z^j$  are the collapse operators taking into account the dephasing on each qubit with rate $\gamma_j$. 

We study the dynamics of this simple two--qubit model and its applicability to the observed dynamics in the Markovian regime. We use the built-in mesolve function in the QuTiP package \cite{qutip_1, qutip_2} to numerically solve for Eq. \ref{eq:lindblad} in the rotating frame to resolve the system's density matrix at various time steps. We initialize the two-qubit $\rho$ in an entangled state consistent with the experiment and set $\gamma_A = 0.024\ \mu\mathrm{s}^{-1}$, consistent with the Ancilla's dephasing time. We then find the best-fit simulation by tuning the value of $\gamma_Q$, modeling the coupling to the environment as a simple Markovian bath.  Figure~\ref{fig:conc_sim} compares these simulation results to the experimental data presented in Fig.~\ref{zeno}. These simulations match the experimental data for all cases except $\gamma = 1.6\ \mu\mathrm{s}^{-1}$, where slight deviations from exponential decay are observed in the data.  This analysis reveals a level of ambiguity about the sharpness of the transition between non-Markovian and Markovian dynamics. Indeed there exists a parameter range (in our case between $\gamma \approx 1$ and $\gamma \approx 3 \ \mu\mathrm{s}^{-1}$) where the non-Markovian measure $\mathcal{N} = 0$, but the GKSL master equation does not exactly match the experimental data. Detailed analysis of this transition region will be the subject of future investigation. 
}

\begin{figure*}[h!]
    \centering
    \includegraphics[width=1.0\textwidth]{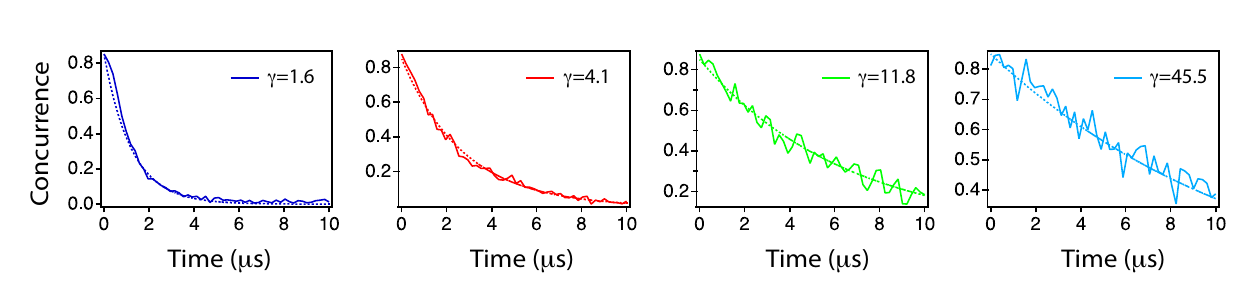}
    \caption{\kater{\textbf{Concurrence evolution of the Qubit--Ancilla system in the Markovian regime.} We compare data for the concurrence to results from the simulation of Eq. \ref{eq:lindblad} for the four cases displayed in Fig.~\ref{zeno}. }}
    \label{fig:conc_sim}
\end{figure*}

\section{Calibration of the Qubit-Environment parametric coupling}

 When the Environment is subject to the additional dissipation channel formed by probing its readout resonator with pseudo-thermal drive, this additional dissipation affects the parametric resonance between the Qubit and Environment. As such, we calibrate the coupling for every value of the noise amplitude.  Figure \ref{fig:para_calib} details this calibration.  The pseudo-thermal drive on the Environment resonator is generated by frequency modulating a monochromatic tone with Gaussian noise from a function generator. The spectrum of the resulting noise is displayed in Fig.~\ref{fig:para_calib}(a).  Figure \ref{fig:para_calib} (b) displays the spectroscopy of the parametric coupling for different noise amplitudes. We prepare the Qubit in the excited state and measure the Environment excitation probability as a function of the parametric drive applied to both the Qubit and the Environment. For appreciable noise amplitude, the resonance is characteristic of a thermal distribution, Fig.~\ref{fig:para_calib}(c). By fitting these resonances, we determine the FWHM linewidth. In Fig.~\ref{fig:para_calib}(d), we compare the additional broadening observed in the parametric resonance (calculated as $8 \pi \times \mathrm{FWHM}$, since both the Qubit and Environment are modulated at their flux sweet-spots), to the Environment dephasing rate measured in Fig.~\ref{transition}(b).  We observe similar scaling of the dephasing with noise amplitude.  Overall, the clean behavior of the parametric coupling, even at very high values of the Environment dephasing, indicates that the parametric drive still functions in the presence of the additional drive on the Environment's resonator.

\begin{figure*}[b!]
    \centering
    \includegraphics[width=\textwidth]{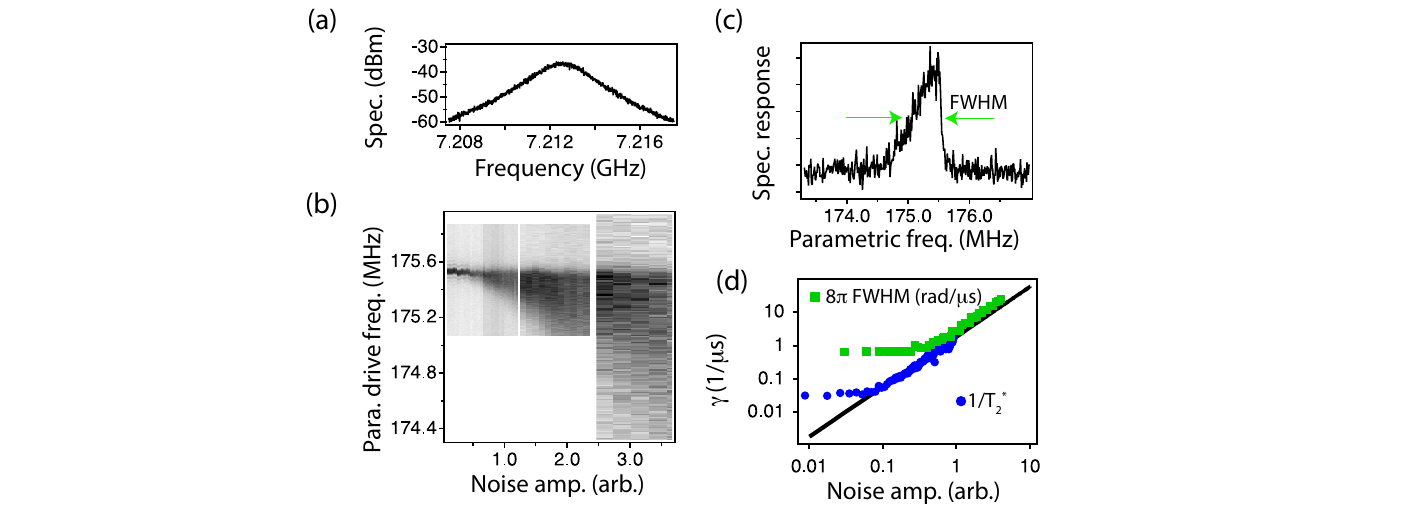}
    \caption{\textbf{Qubit--Environment parametric resonance in the presence of pseudo-thermal noise.} (a) The spectrum of the microwave drive applied at the Environment readout resonator frequency. (b) Spectroscopy of the parametric resonance as given by the Environment excitation probability for different values of the noise amplitude.   (c) The observed spectra are asymmetric as is expected from a thermal photon distribution in the cavity. We determine the FWHM of the resonance as an additional measure of the Environment dephasing rate. (d) The parametric resonance bandwidth (scaled to comparable units for the dephasing rate) versus noise amplitude in comparison to the Environment dephasing rate from Fig.~\ref{transition} (b). This shows that the scaling of the Environment dephasing continues for larger ranges of the noise amplitude than can be probed with Ramsey measurements.}
    \label{fig:para_calib}
\end{figure*}

\section{Experimental Setup} \label{appen:setup}

\begin{figure*}[h!]
    \centering
    \includegraphics[width=0.75\textwidth]{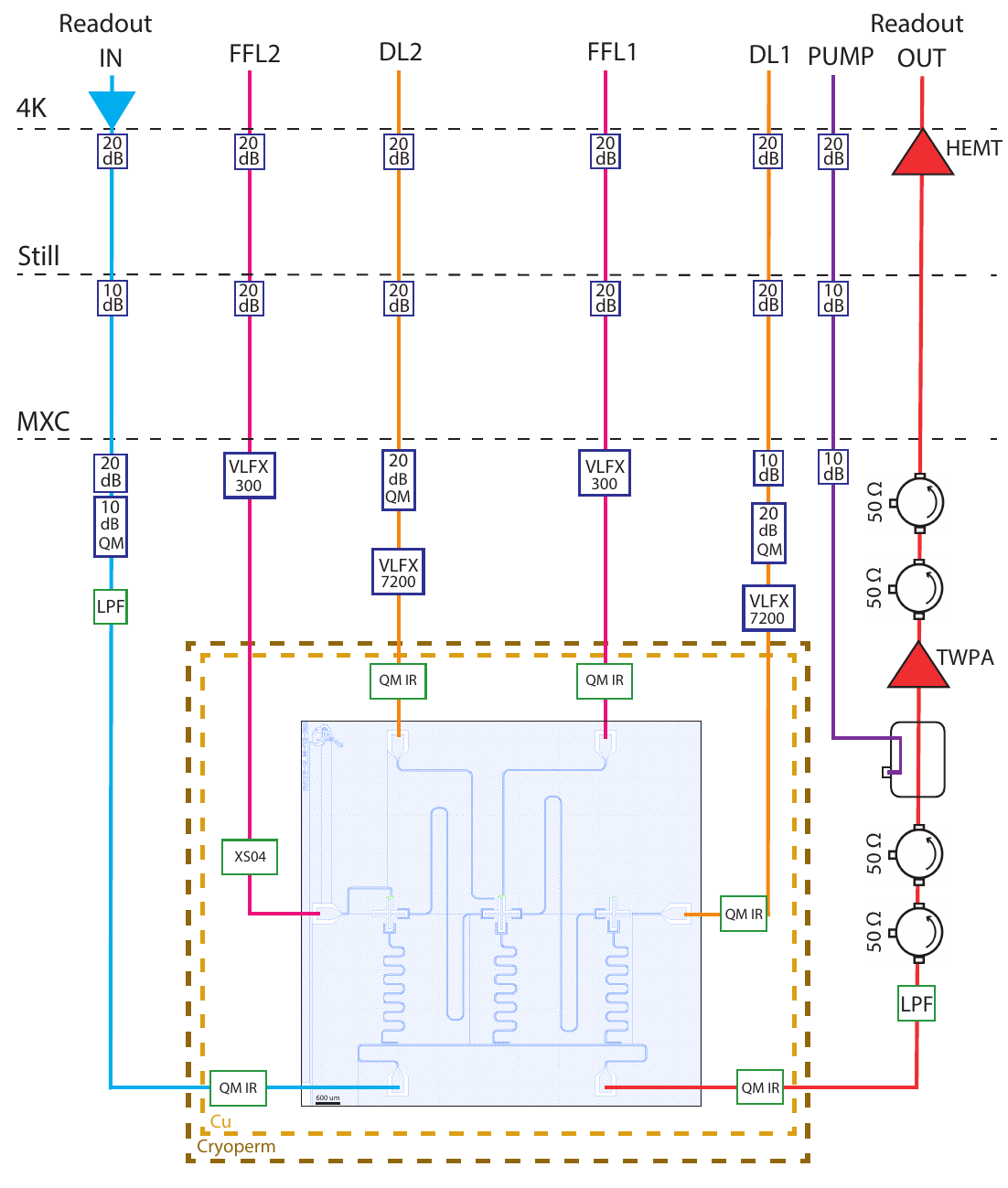}
    \caption{\textbf{Cryogenic setup of the experiment.}}
    \label{fig:fridge}
\end{figure*}

Figure~\ref{fig:fridge} shows the cryogenic setup of the experiment. The device is packaged in a copper box and surrounded by an additional copper can as well as a Cryoperm shielding to protect the device from infrared radiation and external magnetic fields. The device is further thermalized to the mixing chamber stage via a copper plate. The coaxial lines are thermalized via cryogenic microwave attenuators. Note that, the very last attenuators are specific cryogenic attenuators (QMC-CRYOATTF).

For the fast flux lines (FFL), we used a total of $40$ dB attenuation as well as $300$ MHz low-pass filters (MiniCircuits $300$ VLFX) to suppress the high-frequency noise, with a bias-tee at the top of the fridge to apply a DC current to tune the frequency of the qubits. The drive lines DL1 and DL2 have different values of attenuation (70 dB and 60 dB) to account for respective differences in their on-chip coupling. This arrangement of the attenuators allows us to achieve Rabi oscillations as fast as $20$ MHz. In addition, we installed $7.2$ GHz low-pass filters (MiniCircuits $7200$+
VLFX) to mitigate high-frequency noise. Finally, for the readout input line, we added $60$ dB of attenuation with a KNL low-pass filter (LPF) at $8$ GHz (4L250-7344/T12000-O/O). An Eccosorb infrared filter [labeled QM IR (QMC-CRYOIRF-002MF), or XS04 (an equivalent element)] was installed for every single microwave line inside the copper shielding with $> 10$ GHz cutoff frequencies to absorb the infrared radiation. 

To amplify the output signal, we use a high-electron-mobility transistor (HEMT) low-noise amplifier at the 4K stage as well as a traveling-wave parametric amplifier (TWPA) at milliKelvin temperatures with gains of about $40$ dB and $20$ dB, respectively. The TWPA used in this experiment is based on the SNAIL architecture resulting in reversed Kerr phase matching \cite{Ranadive2022} with bandwidths as high as $4$ GHz and noise temperatures of about $300$ mK. The other advantage of this type of TWPA is that it can be pumped at frequencies ${\sim} 2$ GHz away from the range of interest through a directional coupler, which results in minimal interference between the pump and readout signal in the experiment. 

\section{Device design and simulations}\label{appen:sim}

\begin{table*}

\begin{center}
\begin{tabular}{||c | c | c | c | c | c|c|c||} 
\hline 
& \thead{$\omega_\mathrm{q}/2\pi$ (GHz)} & \thead{$|\alpha|/2\pi$ (MHz)} & \thead{$\chi_\mathrm{qc}/2\pi$ (kHz)} & \thead{$\omega_\mathrm{c}/2\pi$ (GHz)} & \thead{$\kappa/2\pi$ (kHz)} & \thead{$T_1\ (\mu\mathrm{s})$} &\thead{$T_2^*\ (\mu\mathrm{s})$} \\ [0.5ex] 
\hline\hline
\thead{Ancilla} & 4.6 [4.2] & 195 [212] & 210 [230] & 7.15 [6.94] & 200 [270] &[32]&[41] \\ [0.5ex] 
\hline
\thead{Qubit} & 5.1 [4.65] & 175 [180] & 210 [250] & 7.3 [7.09] & 200 [206] &[31]&[39] \\ [0.5ex] 
\hline
\thead{Env.} & 5.6 [5.37] & 180 [140] & 200 [265] & 7.47 [7.21] & 200 [170] &[28]&[38]\\[0.5ex] 
 \hline
 
\end{tabular}
\end{center}
\caption{ Simulated [measured] parameters of the device used in the experiment.}
\label{tab:sim}

\end{table*}

The device layout is designed using the Qiskit Metal package \cite{qiskit-metal} by incorporating the prominent Xmon qubit geometry \cite{Barends2013a}. The layout is then imported to the Ansys software in order to perform the finite-element simulations utilizing the eigenmode solver. The simulation results are then imported to the energy-participation quantization package \cite{pyEPR_sw} to extract the qubits' frequencies ($\omega_\mathrm{q}$) and anharmonicities ($\alpha$) as well as the readout resonators' frequencies ($\omega_\mathrm{c}$) and qubit--cavity dispersive shifts ($\chi_\mathrm{qc}$). Moreover, the linewidth of the resonators ($\kappa$) is estimated using the HFSS-driven modal scattering simulations after employing the $3$-dB method \cite{Wisbey2014a} to the scattering transmission profile ($\mathrm{S}_{12}$) from the simulations. Table~\ref{tab:sim} shows the simulated parameters of the device used in the experiment with the measured values written in brackets, indicating an accuracy of 90\% between simulations and measurements. Additionally, the Ancilla--Qubit and Qubit--Environment mediating resonators' frequencies are designed to be $8.0$ and $8.6$ GHz, respectively.

\section{Two-qubit state tomography}\label{appen:2qst}


The density matrix of two qubits can be reconstructed by performing 9 distinct measurements. As an example, if our system is prepared in an arbitrary state in the $\{00, 01, 10, 11\}$ basis as, $\ket{\psi} = a \ket{00} + b \ket{01} + c \ket{10} + d \ket{11} $, the $\braket{ZZ}$ expectation value can be written as, 

\begin{align}
\braket{ZZ} &= \frac{\bra{\psi} \sigma_z \otimes \sigma_z \ket{\psi}}{\langle\psi|\psi\rangle} = \frac{|a|^2 - |b|^2 - |c|^2 + |d|^2}{|a|^2 + |b|^2 + |c|^2 + |d|^2}\nonumber\\ 
&=\frac{\mathrm{pr}^{(z)}(00) - \mathrm{pr}^{(z)}(01) - \mathrm{pr}^{(z)}(10) + \mathrm{pr}^{(z)}(11)}{\rm{pr}^{(z)}(00) + \mathrm{pr}^{(z)}(01) + \mathrm{pr}^{(z)}(10) + \mathrm{pr}^{(z)}(11)}\nonumber,
\end{align}

\noindent where $\sigma_z$ is the Pauli $z$-matrix and $\rm{pr}^{(z)}(00)$ denotes the probability of finding both qubits in their ground states measured along the $Z$ axis. This is achievable by employing a simultaneous heterodyne readout scheme, where we multiplex the readout signal by sending two pulses simultaneously with different intermediate frequencies to track the state of the qubits. A similar approach can be used to measure the rest of the expectation values by applying 9 various combinations of the tomography pulses. Table~\ref{tab:2qst} includes the required rotation pulse combinations and the corresponding expectation values that can be calculated from each measurement, ignoring the trivial $\braket{\mathcal{I}\mathcal{I}}$.

Having all the 16 expectation values allows us to reconstruct the full density matrix following these steps:

\begin{enumerate}

\item Define a lower-triangular matrix as, 

\begin{equation}\label{TMatrix} T = 
\begin{pmatrix}
t_1 & 0 & 0 & 0\\
t_5 + i t_6 & t_2 & 0 & 0\\
t_7 + i t_8 & t_9 + i t_{10} & t_3 & 0\\
t_{11} + i t_{12} & t_{13} + i t_{14} & t_{15} + i t_{16} & t_4
\end{pmatrix}\nonumber,
\end{equation}

\noindent with the density matrix constructed as $\rho = \frac{T^{\dagger} T} {\tr(T^{\dagger} T)}$. This assures the Hermiticity of $\rho$ as well as its trace-normalization.

\item Construct a minimization vector in the form of, $\mathcal{L}_{k} = \left({\tr(\hat{\mathcal{M}}_{k}~\rho) - p_{k}}\right)^2$, where $\hat{\mathcal{M}}_k$ represents the measurement operators (shown in Table~\ref{tab:2qst}) with $p_k$ being the outcome of each measurement and $k$ runs over all the 16 measurement values.

\item Employ the least\_squares function under the scipy.optimize package \cite{SciPy} to minimize the vector defined in step 2 to extract the optimized values for the $T$ matrix elements and reconstruct the density matrix.  

\end{enumerate}

The steps above form the well-known maximum likelihood estimation method, which results in high accuracy in reconstructing the full state of the qubits \cite{James2001}.

\begin{table}
\begin{center}
\begin{tabular}{||c | c | c ||} 
 \hline
 \thead{Rotation} & Measurement operator & Expectation values  \\ [0.5ex] 
\hline\hline
 \thead{$R_Y^{\pi/2} \otimes R_Y^{\pi/2}$} & $XX$ & $\braket{IX}, \braket{XI}, \braket{XX}$  \\ [0.5ex] 
 \hline
 \thead{$R_Y^{\pi/2} \otimes R_X^{-\pi/2}$} & $XY$ & $\braket{IY}, \braket{XI}, \braket{XY}$  \\[0.5ex] 
 \hline
 \thead{$R_Y^{\pi/2} \otimes \mathcal{I}$} & $XZ$ & $\braket{IZ}, \braket{XI}, \braket{XZ}$ \\[0.5ex] 
  \hline
  \thead{$R_X^{-\pi/2} \otimes R_Y^{\pi/2}$} & $YX$ & $\braket{IX}, \braket{YI}, \braket{YX}$  \\[0.5ex] 
\hline
 \thead{$R_X^{-\pi/2} \otimes R_X^{-\pi/2}$} & $YY$ & $\braket{IY}, \braket{YI}, \braket{YY}$  \\[0.5ex] 
\hline
 \thead{$R_X^{-\pi/2} \otimes \mathcal{I}$} & $YZ$ & $\braket{IZ}, \braket{YI}, \braket{YZ}$  \\[0.5ex] 
\hline
 \thead{$\mathcal{I} \otimes R_Y^{\pi/2}$} & $ZX$ & $\braket{IX}, \braket{ZI}, \braket{ZX}$  \\[0.5ex] 
 \hline
 \thead{$\mathcal{I} \otimes R_X^{-\pi/2}$} & $ZY$ & $\braket{IY}, \braket{ZI}, \braket{ZY}$  \\[0.5ex] 
 \hline
 \thead{$\mathcal{I} \otimes \mathcal{I}$} & $ZZ$ & $\braket{IZ}, \braket{ZI}, \braket{ZZ}$ \\  [0.5ex] 
\hline
\end{tabular}
\end{center}
\caption{\bf Two-qubit tomography measurement operators and their corresponding 16 expectation values.}
\label{tab:2qst}
\end{table}
	

\end{widetext}

\end{document}